\documentclass[aps,12pt,superscriptaddress,preprint]{revtex4}
\usepackage{graphicx}
\usepackage{amsmath}
\usepackage{hyperref}
\usepackage{bm}
\usepackage[dvips]{color}

\newcommand{\fwi} {Helmholtz-Zentrum Dresden-Rossendorf, Institute of Ion Beam Physics and Materials Research, Bautzner Landstr. 400, 01328 Dresden, Germany}
\newcommand{\fwk} {Helmholtz-Zentrum Dresden-Rossendorf, Institute of Radiation Physics, Bautzner Landstr. 400, 01328 Dresden, Germany}
\newcommand{\tud} {Technische Universit\"{a}t Dresden, 01062 Dresden, Germany}

\newcommand{\jena} {Institut f\"{u}r Festk\"{o}rperphysik, Friedrich-Schiller Universit\"{a}t Jena, 07743 Jena, Germany}
\newcommand{\iop} {Research \& Development Center for Functional Crystals, Beijing National Laboratory for Condensed Matter Physics, Institute of Physics, Chinese Academy of Sciences, Beijing 100190, China}
\newcommand{\TUC} {Faculty of Science, Technische Universit\"at Chemnitz, 09107 Chemnitz, Germany}
\newcommand{\cfaed} {Center for Advancing Electronics Dresden,  Technische Universit\"at Dresden, 01314 Dresden, Germany}
\newcommand{\issp} {Key Laboratory of Materials Physics, Institute of Solid State Physics, Chinese Academy of Sciences, Hefei 230031, People's Republic of China}
\newcommand{\hldhefei} {High Magnetic Field Laboratory, Hefei Institutes of Physical Science, Chinese Academy of Sciences, Hefei 230031, People's Republic of China}
\newcommand{\hzb} {Helmholtz-Zentrum Berlin f\"{u}r Materialien und Energie GmbH, Abteilung F-A1, Hahn-Meitner-Platz 1, 14109 Berlin, Germany}

\begin{document}

%\title{Defect induced paramagnetism and ferromagnetism in bulk SiC}
%\title{Defect-induced magnetism in SiC: Why is the ferromagnetic signal weak?}
%\title{Influence of defects on the paramagnetism and ferromagnetism in bulk SiC}
\title{Defect induced magnetism in bulk SiC: interplay between ferromagnetism and paramagnetism}

\date{\today}

\author{Yutian~Wang}
\address{\fwi}
\address{\tud}
\author{Yu~Liu}
\address{\fwi}
\address{\iop}
\author{Elke~Wendler}
\address{\jena}
\author{Rene~Huebner}
\address{\fwi}
\author{Wolfgang~Anwand}
\address{\fwk}
\author{Gang~Wang}
\address{\iop}
\author{Xuliang~Chen}
\address{\issp}
\author{Wei~Tong}
\address{\hldhefei}
\author{Zhaorong~Yang}
\address{\issp}
\author{Frans~Munnik}
\address{\fwi}
\author{Gregor~Bukalis}
\address{\hzb}
\author{Xiaolong~Chen}
\address{\iop}
\author{Sibylle~Gemming}
\address{\fwi}
\address{\TUC}
\address{\cfaed}
\author{Manfred~Helm}
\address{\fwi}
\address{\tud}
\address{\cfaed}
\author{Shengqiang~Zhou}
\email[Electronic address: ]{s.zhou@hzdr.de}
\address{\fwi}

\begin{abstract}

Defect-induced ferromagnetism has triggered a lot of
investigations and controversies. The major issue is that the
induced ferromagnetic signal is so weak that it can sufficiently
be accounted for by trace contamination. To resolve this issue, we
studied the variation of the magnetic properties of SiC after
neutron irradiation with fluence covering four orders of
magnitude. A large paramagnetic component has been induced and
scales up with defect concentration, which can be well accounted
for by uncoupled divacancies. However, the ferromagnetic
contribution is still weak and only appears in the low fluence
range of neutrons or after annealing treatments. First-principles
calculations hint towards a mutually exclusive role of the
concentration of defects: a higher defect-concentration favors a
larger magnetic interaction at the expense of spin polarization.
Combining both experimental and first-principles calculation
results, the defect-induced ferromagnetism can probably be
understood as a local effect which cannot be scaled up with the
volume.

\end{abstract}
\maketitle

\section{Introduction}
The magic of magnetism was disclosed in the early 20th century
with the development of quantum mechanics. The
Heisenberg model has since then been extremely successful to
understand magnetism and magnetic materials
\cite{coeybook}. All previously identified ferromagnetic bulk
materials contain elements with partially filled 3\emph{d} (4\emph{d}) or 4\emph{f} (5\emph{f}) shells.
A fundamental question is whether materials containing only \emph{s} or
\emph{p} electrons can be ferromagnetic. For nearly two decades,
there have been various theoretical and experimental studies
devoted to clarifying this question \cite{coey2002ferromagnetism,PhysRevLett.91.227201,coey2004magnetism,
PhysRevLett.95.097201,osorio2006magnetism,yoon06,PhysRevLett.99.107201,PhysRevB.79.113201,Cervenka2009,droghetti2010polaronic,wu2010magnetism,6ISI:000223085400033,straumal2009magnetization,PhysRevB.80.035331,dev:117204,Roever2011,PhysRevLett.104.137201,Jcoey1367,slipukhina2011ferromagnetic}.
It turns out that materials with completely filled 3\emph{d} or
4\emph{f} shells or with only \emph{s} or \emph{p} electrons can
be ferromagnetic when they contain defects.
Among those materials, graphite/graphene and oxides attract
particular attention due to experimental evidence reported by
various groups \cite{PhysRevLett.91.227201,
PhysRevLett.95.097201,Xia2008,Cervenka2009,Shukla20121817,han2003observation,PhysRevB.72.224424,PhysRevB.75.075426,18ISI:000302630100016,qin2014strong}.
However, the experimentally measured ferromagnetism remains a
weak signal slightly above the detection limit of sensitive
SQUID magnetometry
\cite{PhysRevB.79.113201,58PhysRevB.81.214404,59PhysRevB.85.144406,Liu2011,Lilin,Roever2011}.
The very weak magnetization not only limits the practical
applications, but also raises questions on the fundamentals of
defect induced ferromagnetism. On the one hand, measurement
artifacts in SQUID magnetometry may occur: improper mounting of
samples and wrong use of sample holders can easily generate
ferromagnetic like signal
\cite{abraham:252502,sawicki2011sensitive,SQUID}. On the other hand,
the debate over the purity of graphite and oxide substrates
continues in parallel
\cite{Esquinazi2010a,Sepioni2012,0295-5075-98-5-57006,0295-5075-98-5-57007,spemann2013trace}.
Pristine graphite and oxides substrates are often ferromagnetic
due to different contaminations or due to intrinsic defects
\cite{Sepioni2012,golmar:262503,PhysRevB.81.214414,ISI:000267045400001},
which hamper the interpretation of the observed ferromagnetism.
Indeed, Nair et al. reported that there is only paramagnetism in
graphene after introducing adatoms or defects \cite{Nair}.

Therefore, the understanding of defect-induced
ferromagnetism in materials without partially filled 3\emph{d} or
4\emph{f} electron shells is far from satisfactory. It is rather calling for an
investigation in which the following requirements should be
fulfilled:

\begin{itemize}
    \item The pristine materials should be well controlled with the highest purity grade possible.
    \item The materials should be supplied in a large quantity with identical properties, such that one can perform a series of experiments using identical specimens.
    \item The materials should be free of elements with 3\emph{d} or 4\emph{f} electrons.
    \item The induced effect should exist in a large volume to measure a large enough magnetic signal.
\end{itemize}

The last point was firstly proposed by Coey et al., who suggested to measure a bulk graphite-nodule to gain a large ferromagnetic signal \cite{coey2002ferromagnetism}. However, minor amounts of magnetite, kamacite, etc, also appear in their graphite samples \cite{coey2002ferromagnetism}. These secondary phases are responsible for about two-thirds of the observed magnetization and the remaining one-thirds is attributed to the graphite-nodule.

After screening by these facts, Si and SiC are the best
candidates. Defect-induced ferromagnetism was revealed in SiC
after neutron and ion irradiation
\cite{Liu2011,Lilin}. Recently, it was suggested
that the \emph{p} electrons from the carbon atoms are mainly
responsible for the long-range ferromagnetic coupling in SiC after
ion irradiation \cite{WangXMCD}, which is similar to the
observation in graphite \cite{1367-2630-12-12-123012}. In this
work, we start with 4H-SiC, high purity semi-insulating SiC.
Neutron irradiation was used to introduce defects in the whole
sample volume over a large range of defect concentrations. The
large ``magnetic volume'' and well controlled pristine materials
allow for reliable interpretation for the following experimental
observations: (1) paramagnetism increases with neutron
fluence and saturates at the end; (2) ferromagnetism only occurs
in a narrow fluence range although it is weak; (3) thermal
annealing paramagnetic defective SiC can induce weak
ferromagnetism. Together with density-functional theory (DFT), we show that there is
an intrinsic limit to increase the ferromagnetic contribution to the magnetization.

\section{Experiment}
Commercial semi-insulating 4H-SiC single crystal wafers (Cree) were
used for our experiment. Particle induced X-ray emission (PIXE) was
performed to check magnetic impurities in the pristine sample. The
concentrations of magnetic impurities (Fe, Co and Ni) are proved
to be below the detection limits (shown later). Neutron
irradiation was performed at chambers DBVK and DBVR at the BER II
reactor at Helmholtz-Zentrum Berlin at a temperature less than
50 $^{\circ}$C. The fluxes of DBVK are 1.08$\times$10$^{14}$
cm$^{-2}$s$^{-1}$ of thermal neutrons, 7.04$\times$10$^{12}$
cm$^{-2}$s$^{-1}$ of epithermal neutrons, and
5.80$\times$10$^{13}$ cm$^{-2}$s$^{-1}$ of fast neutrons,
respectively and those of DBVR are 1.50$\times$10$^{14}$
cm$^{-2}$s$^{-1}$, 6.80$\times$10$^{12}$ cm$^{-2}$s$^{-1}$,
5.10$\times$10$^{13}$ cm$^{-2}$s$^{-1}$, correspondingly. Comparing with epithermal or fast neutrons, thermal neutrons only produce negligible displacement. Therefore, only epithermal and fast neutrons are accounted for in the fluence calculation \cite{wendler2012damage}. The neutron fluence spanned a large range from
3.28$\times$10$^{16}$ cm$^{-2}$ to 3.50$\times$10$^{19}$ cm$^{-2}$
covering all possibilities from light damage to near
amorphization. It is worthy to note that the applied neutron
fluence covers the range used in Ref. \cite{Liu2011} and is extended to much higher
fluence values by two orders of magnitude. For structural
characterization, $\mu$-Raman spectroscopy has been performed by
using a Nd:YAG Laser with 532 nm wavelength in the
scattering geometry using a liquid nitrogen cooled charge-coupled
device camera. Positron annihilation Doppler broadening spectroscopy (DBS) was applied to clarify the nature of defects using the mono-energetic slow position beam "SPONSOR" \cite{anwand2012design}. High-resolution transmission electron microscopy (HR-TEM) investigations were done with an image-corrected Titan 80-300 microscope (FEI) operated at an accelerating voltage of 300 keV. The magnetic properties are measured by a superconducting quantum interference device magnetic property
measurement system and vibrating sample magnetometers (SQUID-MPMS
and SOUID-VSM, Quantum Design). The electron spin resonance (ESR) spectroscopy was performed at 9.4 GHz using a Bruker spectrometer (Bruker ELEXSYS E500).

\section{Results}

\subsection{Virgin vs. irradiated SiC}
Before we start to describe the detailed results, we first present the key facts for the virgin SiC substrates employed in our investigation.

\begin{figure} \center
\includegraphics[scale=0.5]{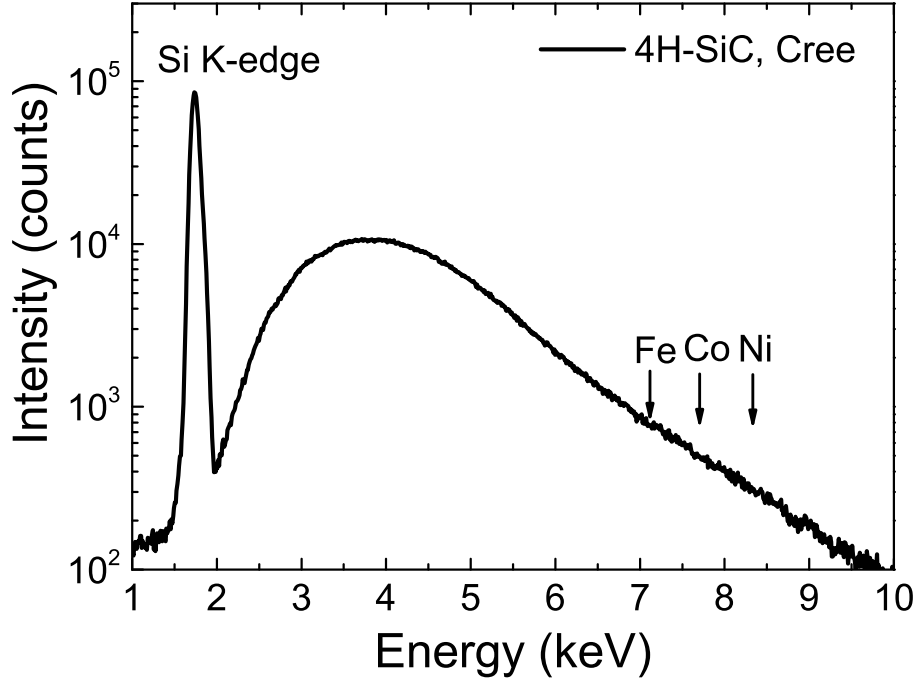}
\caption{PIXE spectrum for the 4H-SiC wafer by a broad proton beam. Within the detection limit, no Fe, Co or Ni contamination is observed.} \label{4HSiC_Cree_PIXE}
\end{figure}

To check the trace presence of transition metals (Fe, Co and Ni, etc.) in our SiC substrates, we have performed PIXE using 2 MeV protons with a broad beam of around 1 mm$^2$, as shown in Figure \ref{4HSiC_Cree_PIXE}. PIXE is a sensitive method to detect trace impurities in bulk volume without significant structural destruction \cite{Esquinazi2010a,spemann2013trace}. In the spectrum, the sharp peak is the Si K-line X-ray emission. The broad bump is due to the secondary electron Bremsstrahlung background. For commercially available, purest graphite, there is detectable transition metal contamination (mainly Ti, V, Fe, Ni) as revealed by PIXE \cite{spemann2013trace}. In sharp contrast, the transition metal impurity in the SiC we used for this study, if there is, is below the detection limit of around 1 ppm (Fig. \ref{4HSiC_Cree_PIXE}).

\begin{figure} \center
\includegraphics[scale=0.5]{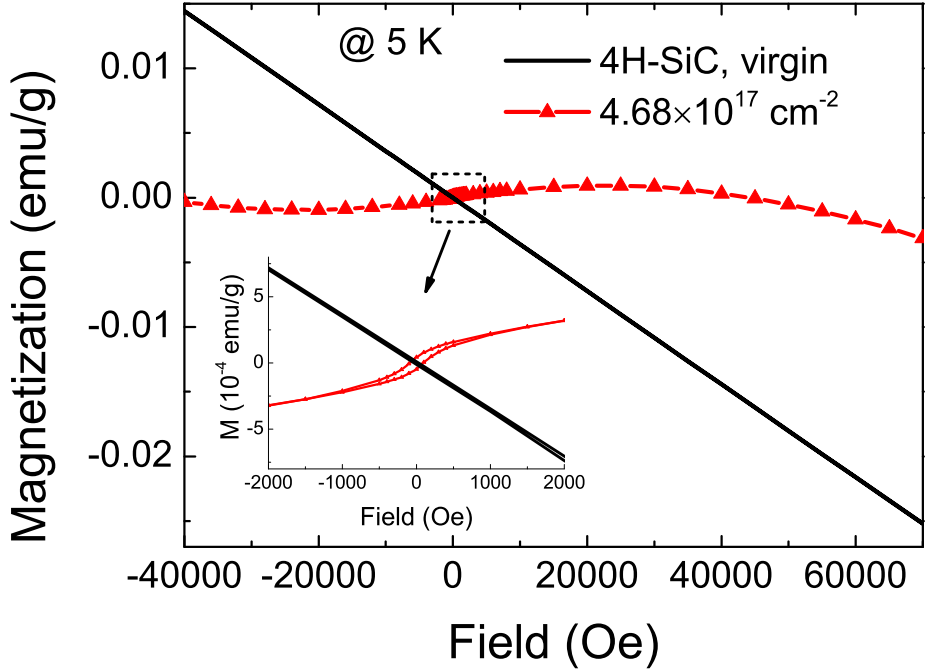}
\caption{Comparison of the magnetic properties: virgin vs. neutron irradiated SiC. Inset: the zoom of the low field part. Neutron irradiation induces significant, unambiguous magnetic variation in SiC.} \label{nSiC_virgin_M}
\end{figure}

Figure \ref{nSiC_virgin_M} shows a comparison of the magnetic properties measured at 5 K between the virgin SiC and a specimen irradiated with neutrons with a relative low fluence value of 4.68$\times$10$^{17}$ cm$^{-2}$. The results are presented without any background subtraction but only normalized to the mass of the samples. The difference between the virgin and irradiated SiC is obviously not trivial: the virgin SiC shows purely diamagnetic behavior, while the irradiated one shows a paramagnetic component as revealed by the large deviation from the linear dependence on the magnetic field. The inset of Fig. \ref{nSiC_virgin_M} shows the zoom of the low field part. Besides the paramagnetic component induced by    neutron irradiation, a small ferromagnetic contribution also shows up with its saturation magnetization around 1\% of the paramagnetic signal. The defect-induced paramagnetism and ferromagnetism are exactly the key facts we are going to discuss in this manuscript.

\subsection{Structural properties}
\subsubsection{Raman spectroscopy}
\begin{figure} \center
\includegraphics[scale=0.3]{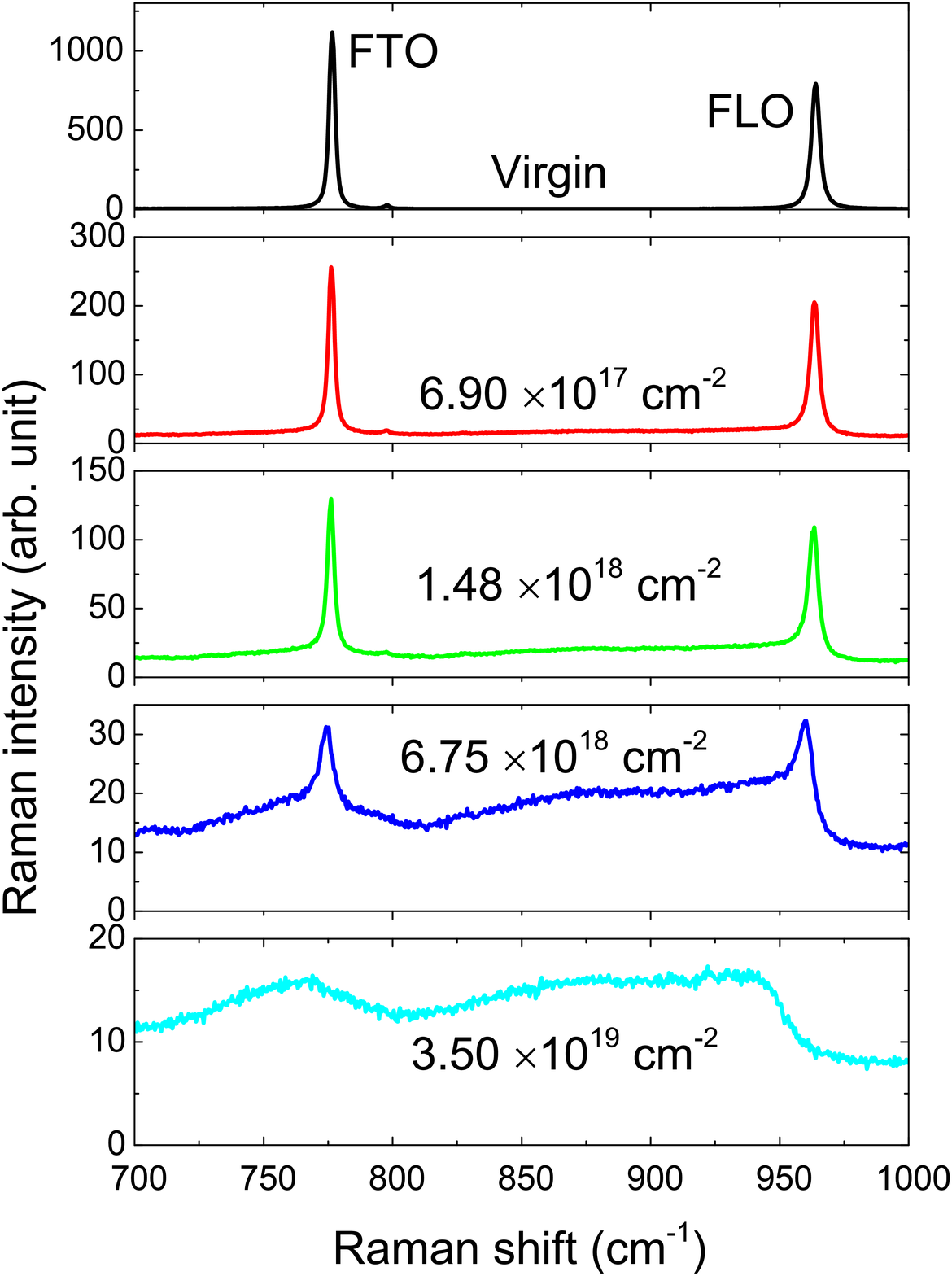}
\caption{Raman spectra for virgin and neutron irradiated 4H-SiC single crystals. The folded transverse optic (FTO) and longitudinal optic (FLO) modes are identified. With increasing neutron fluence the Raman scattering intensity is decreased and the peak is broadened.} \label{FigRaman_nSiC}
\end{figure}

The damage to the crystallinity of SiC upon neutron irradiation is
verified by Raman spectroscopy. Figure \ref{FigRaman_nSiC}
exemplarily shows the Raman spectra for the virgin sample and the
samples irradiated with different neutron fluence values as indicated in
the figure. The typical Raman scattering modes for 4H-SiC are
resolved: folded transverse optic (FTO) and longitudinal optic (FLO) modes \cite{burton1999first}. Note the different scale in the y-axis: with increasing
fluence, the intensity of the Raman scattering modes is
dramatically reduced, as observed in Refs.
\cite{Liu2011,Lilin}. This behaviour
directly reflects the increasing disorder of the crystalline
material. For the sample with largest fluence, the Raman modes are much broad and
hardly resolved due to the large amount of defects \cite{Weber199862,wendler2012damage}. However, after annealing at 900 $^{\circ}$C for 15 min the crystalline order can be substantially recovered.

\begin{figure} \center
\includegraphics[scale=0.5]{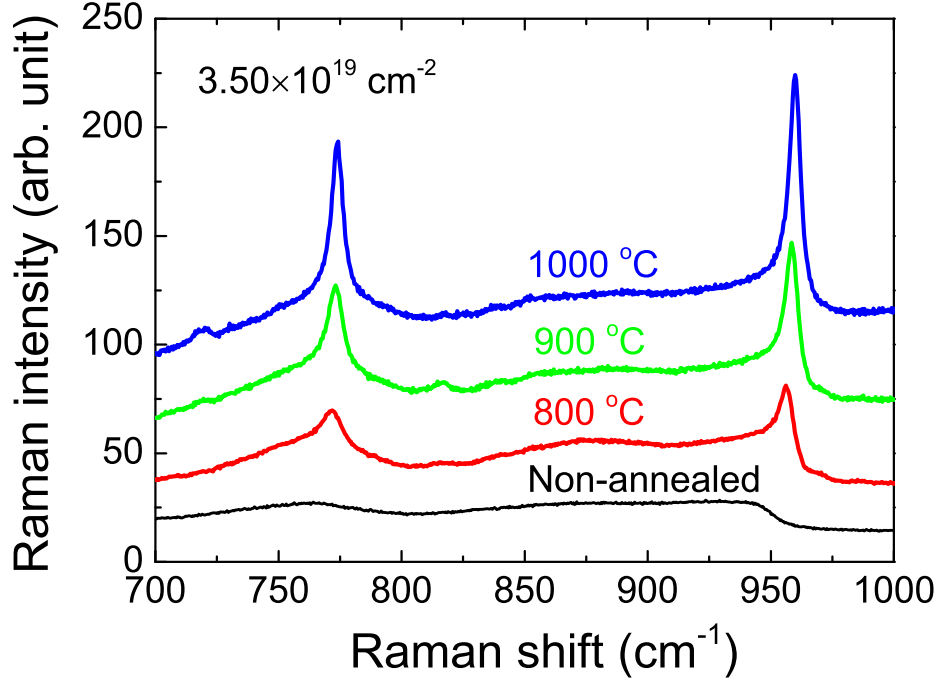}
\caption{Raman spectra for neutron irradiated and thermally annealed 4H-SiC single crystals. The annealing was performed in N$_{2}$ atmosphere 15 min at the temperature indicated. With increasing annealing temperature, the structural damage has been healed.} \label{Raman_annealing}
\end{figure}

Figure \ref{Raman_annealing} shows the Raman spectra of an irradiated SiC (fluence: 3.50$\times$10$^{19}$ cm$^{-2}$) after annealing at different temperature. The annealing was performed in N$_{2}$ atmosphere at 800, 900 and 1000 $^{\circ}$C for 15 min. With increasing annealing temperature, the peaks of the SiC Raman modes become sharper and their intensity gradually increases. The increase of the peak intensity is related to the recovery of the crystalline lattice with annealing. However, a complete recovery of the lattice is not achieved. It has been shown that after annealing at 1450 $^{\circ}$C for 10 min the Raman peaks are comparable to the pristine SiC \cite{bohn1987recrystallization}.

\subsubsection{Positron annihilation spectroscopy}
Positron annihilation Doppler broadening spectroscopy (DBS) is an excellent technique to detect open volume defects from clusters consisting of several vacancies down to a mono-vacancy \cite{anwand2012design}. The positron in a crystal lattice is strongly subjected to repulsion from the positive atom core. Because of the locally reduced atomic density inside the open volume defects, positrons have a high probability to be trapped and to annihilate with electrons in these defects by the emission of two 511 keV photons. The Doppler broadening of the 511 keV annihilation line is mainly caused by the momentum of the electron due to the very low momentum of the thermalized positron. The Doppler broadening parameter $S$, obtained from the 511 keV annihilation line, reflects the fraction of positrons, annihilating with electrons of low momentum (valance electrons). In this study, the $S$ parameter is defined as the ratio of the counts from the central part of the annihilation peak (here 510.17 keV - 511.83 keV) to the total number of counts in the whole peak (498 keV - 524 keV). Therefore, the $S$ parameter is a measure for the open volume in the material. It increases with increasing size of the particular open volume defects.

\begin{figure} \center
\includegraphics[scale=0.5]{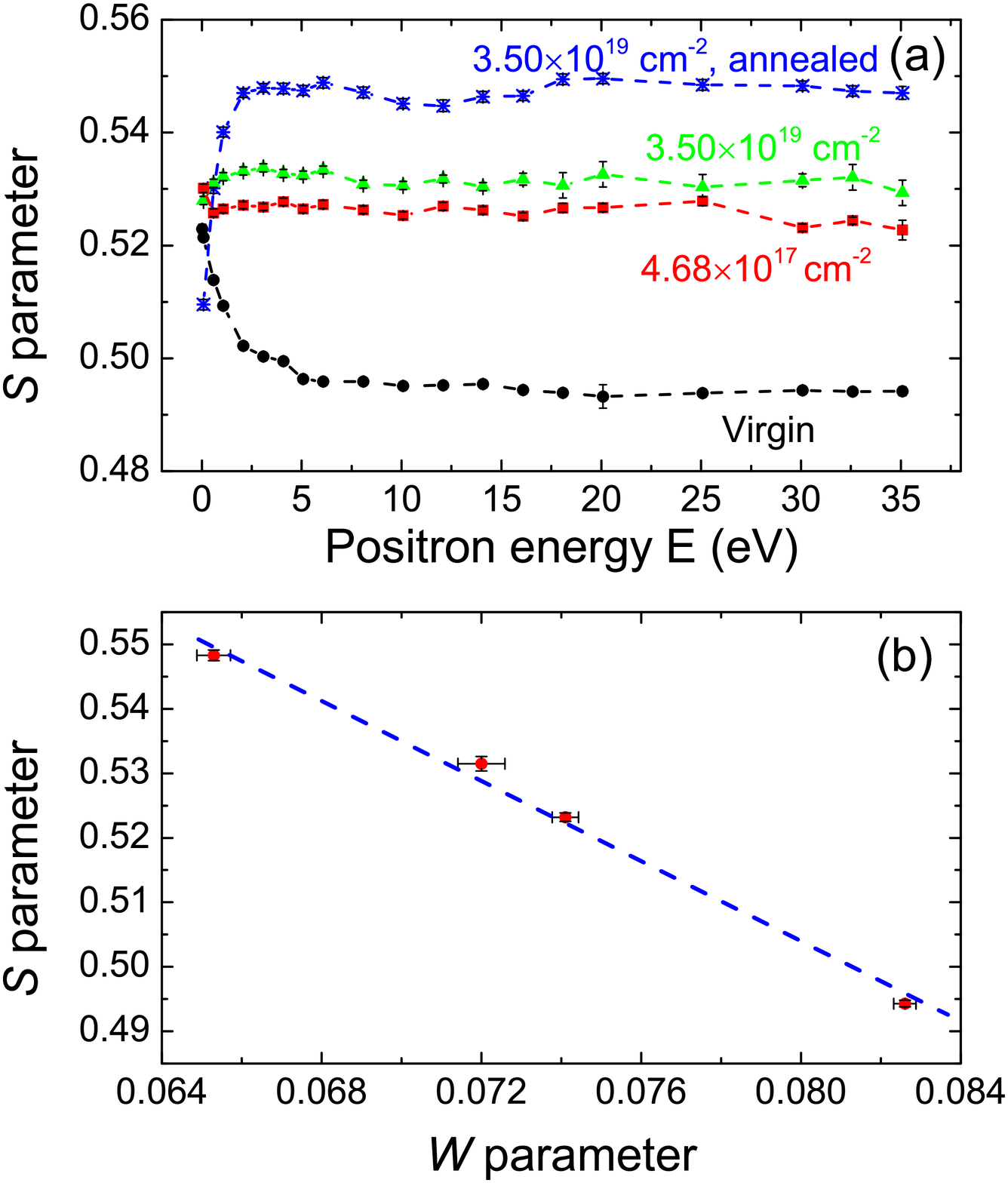}
\caption{(a) Mean $S$ parameter vs. incident positron energy for virgin, neutron irradiated (431-50 with fluence of 3.50$\times$10$^{19}$ cm$^{-2}$ and 431-56 with fluence of 3.50$\times$10$^{19}$ cm$^{-2}$) and thermally annealed (431-56) 4H-SiC single crystals. The annealing was performed in N$_{2}$ atmosphere at 900 $^{\circ}$C for 15 min. (b) $S-W$ plot of DBS results measured on different SiC crystals shown in (a).} \label{pas}
\end{figure}

According to our previous positron annihilation experiments \cite{Liu2011,Lilin}, divacancies have been identified in neutron or ion irradiated SiC. Figure \ref{pas}(a) shows the measured $S$ parameters versus the incident positron energy. The plateau of the $S$ parameters of the irradiated samples above a positron energy of 2 keV corresponds to positron annihilation in the defects created by neutron irradiation. Defects are homogenously distributed along the depth as expected for neutron irradiation. The $S$ parameters of irradiated samples are larger than that of the virgin SiC. $S$ does not increase significantly with increasing neutron fluence, but does increase after annealing. Figure \ref{pas}(b) displays the $S-W$ plot for different samples. A straight line is obtained in the $S-W$ representation for different SiC samples. It shows that the same type of defects, i.e., open volume damage \cite{anwand2002vacancy,brauer2006defects}, exist in different SiC samples. A relation between the ratio $S_{defect}$/$S_{bulk}$ and the number of agglomerated V$_{Si}$V$_{C}$ divacancies in 6H-SiC is published as a scaling curve in Ref. \onlinecite{anwand2002vacancy}. As a rough estimation, the post-irradiation annealing at 900 $^{\circ}$C led to a defect agglomeration up to a size of 8 V$_{Si}$V$_{C}$ divacancies.

One has to note that PAS is only sensitive to negatively charged or neutral open volume defects. The $S$ parameter increases with the size of the open volume defects, but not necessarily with the amount of defects. Raman and magnetization are measurements of total amount of defects. This explains why the $S$ parameter increases after annealing, indicating coalescence of voids to larger but fewer ones, but the Raman and magnetization measurements show the decrease of amount of defects.

\subsubsection{High-resolution transmission electron microscopy}

With the aim to visualize the defects, we performed high resolution transmission electron microscopy on selected samples: a pristine 4H-SiC, 431-56 (irradiated with neutron at the fluence of 3.50$\times$10$^{19}$ cm$^{-2}$) and 431-56 after annealing at 900 $^{\circ}$C for 15 min. Figure \ref{TEM} displays corresponding cross-sectional high-resolution transmission electron micrographs taken in [100] zone axis geometry. It is observed that the 4H-SiC stacking sequence (abcb) perpendicular to the surface has been kept well during neutron irradiation up to the applied fluence. There are no observable contrast differences between the samples, even after annealing. This finding is due to the fact that neutron irradiation (in relatively low fluence regime compared with literature) mainly creates point defects. According to the magnetization results shown later, the average defect concentration for the sample with the largest fluence is below 0.1\%. Furthermore, it should be kept in mind, that classical TEM specimen preparation by sawing, grinding, polishing, dimpling, and final Ar ion milling may also lead to the generation of point defects, resulting in minor contrast variations in the HR-TEM micrographs compared to a completely defect-free single crystalline 4H-SiC specimen. Therefore, TEM seems to be not the right technique to clearly visualize the defects in the relatively low fluence regime. As shown in refs. \onlinecite{Weber199862,yano1990high,leclerc2008evolution,kondo2008microstructural}, defects appear to be visible by TEM when the neutron fluence is above 10$^{21}$ cm$^{-2}$, i.e. 100 times larger than in our case.

\begin{figure*} \center
\centering\fbox{ \begin{minipage}{3in} \hfill\vspace{3in} \end{minipage} } - See more at the original publication Phys. Rev. B 92, 174409 (2015) due to the size limit.
\caption{Cross-sectional HR-TEM micrographs of 4H--SiC single crystals: (a) Pristine SiC (b) 431-56 (irradiated with neutron fluence of 3.50$\times$10$^{19}$ cm$^{-2}$) and (c) 431-56 after annealing at T = 900 $^{\circ}$C.} \label{TEM}
\end{figure*}

\subsection{Paramagnetism}

\begin{figure} \center
\includegraphics[scale=0.35]{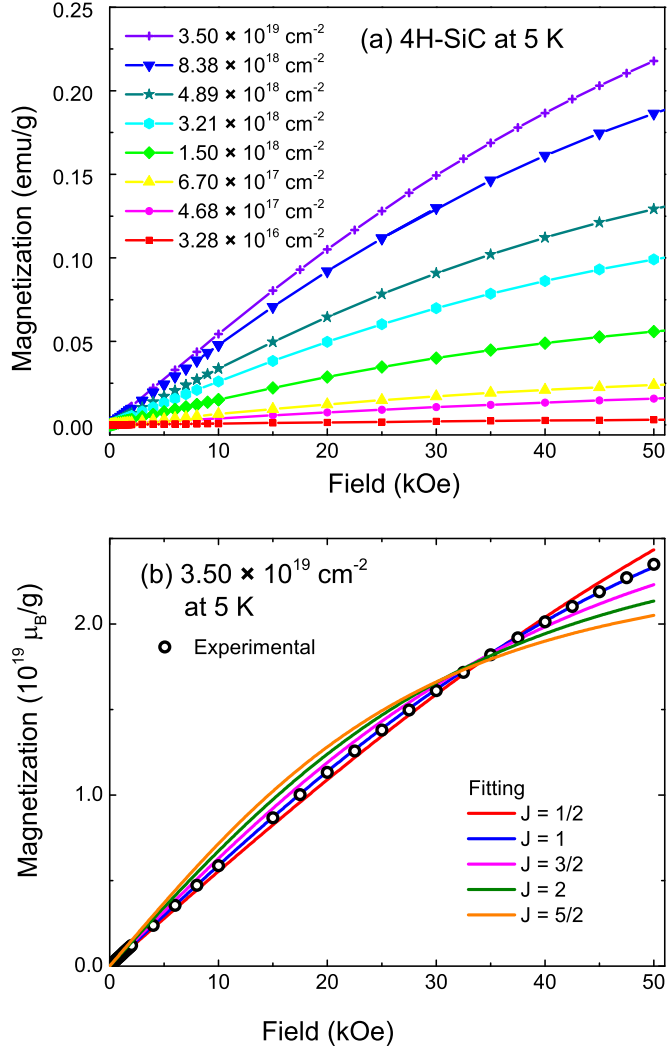}
\caption{(a) Magnetization of all irradiated samples measured at
5 K as a function of field [\emph{(M(H)}], (b) Fits of the magnetization measured at 5 K for the sample with a fluence of 3.50$\times$10$^{19}$ cm$^{-2}$ using
the Brillouin function with different values
of \emph{J}.} \label{figMH_nSiC}
\end{figure}

Figure \ref{figMH_nSiC}(a) presents the magnetization
[\emph{(M(H)}] of neutron irradiated SiC after subtracting the
diamagnetic background which is measured for the pristine sample
at 300 K. It is clearly seen that all samples exhibit a
paramagnetic component. The magnetization increases with
increasing neutron fluence. The paramagnetism can be accurately
described by Brillouin function:

\begin{equation}\label{Brillouin}
M(\alpha)=NJ{\mu_B}g[\dfrac{2J+1}{2J}coth(\dfrac{2J+1}{2J}{\alpha})-\dfrac{1}{2J}coth(\dfrac{\alpha}{2J})]
\end{equation}

where the $g$ factor is about 2 obtained from electron spin
resonance measurement (shown later), $\mu_B$ is the Bohr magneton, $\alpha=gJ\mu_BH/k_BT$, $k_B$ is Boltzmann constant and \emph{N} is the density of spins.
As exemplarily shown in Fig. \ref{figMH_nSiC}(b), the data can be fitted nicely by
\emph{J} = 1. Moreover, a value of \emph{J} larger or smaller than 1 results in a pronounced
deviation from the experimental data. This fitting allows us to conclude
that the observed paramagnetism is due to a single species
with moment $\mu=gJ\mu_B=2 \mu_B$. This conclusion is in excellent agreement with our first-principles calculation (shown later): the induced magnetism is due to V$_{Si}$V$_{C}$ divacancy carrying a magnetic moment of 2 $\mu_B$. As a comparison, the induced magnetism in graphite is generally attributed to single vacancy \cite{PhysRevLett.93.187202,Ugeda2010a,yazyev2010emergence,wang2014defect}. In ZnO, the origin of the local moment is more complicated. Both Zn and O vacancies have been suggested to be responsible for the induced magnetism \cite{PhysRevLett.104.137201,chanier2008magnetic,chan2009electronic,guglieri2014evidence}. On the other hand, oxygen atoms on the polar ZnO surface can be spin polarized and undergo a long-range magnetic interaction \cite{fischer2011room,hernando2011revisiting,ye2012spin}.

\begin{figure} \center
\includegraphics[scale=0.4]{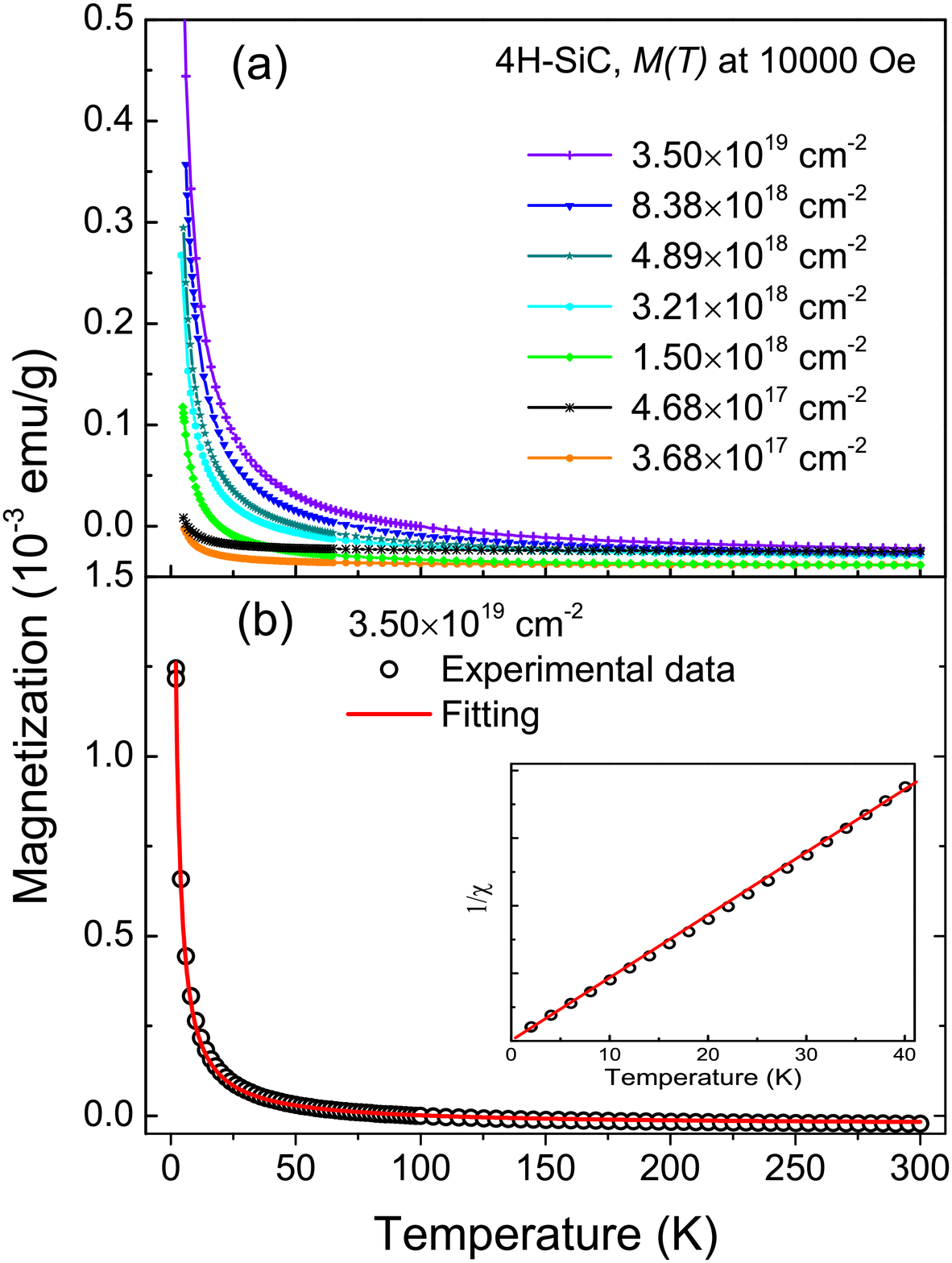}
\caption{(a) The magnetic moment of all irradiated samples measured at
10000 Oe as a function of temperature [\emph{M(T)}]. (b) For the sample irradiated with the largest fluence: the black symbols are experimental data and the red solid curve is the fitting result by equation (2).} \label{FigMT_nSiC}
\end{figure}

Figure \ref{FigMT_nSiC} shows the temperature dependent
magnetization [\emph{M(T)}] for samples with different neutron
fluences under an applied field of 10000 Oe. As expected, one contribution represents the diamagnetic background, which is essentially temperature independent and dominates at higher
temperature. The Curie-like paramagnetic component shows a strong
temperature and neutron fluence dependence. In a self-consistence, the \emph{M(T)} curves can be well fitted according to the Curie law (Eq. \ref{Curie_law}) by using the same \emph{J} and \emph{N} obtained from the corresponding \emph{M(H)}
curve fitting: $J$ = 1 and $N$ = 1.66$\times$10$^{19}$/g for the sample shown in Fig. \ref{figMH_nSiC}(b) and Fig. \ref{FigMT_nSiC}(b).

\begin{equation}\label{Curie_law}
    \chi=\frac{M}{H}=N\frac{J(J+1)(g\mu_B)^{2}}{3k_{B}T}
\end{equation}

\begin{figure} \center
\includegraphics[scale=0.4]{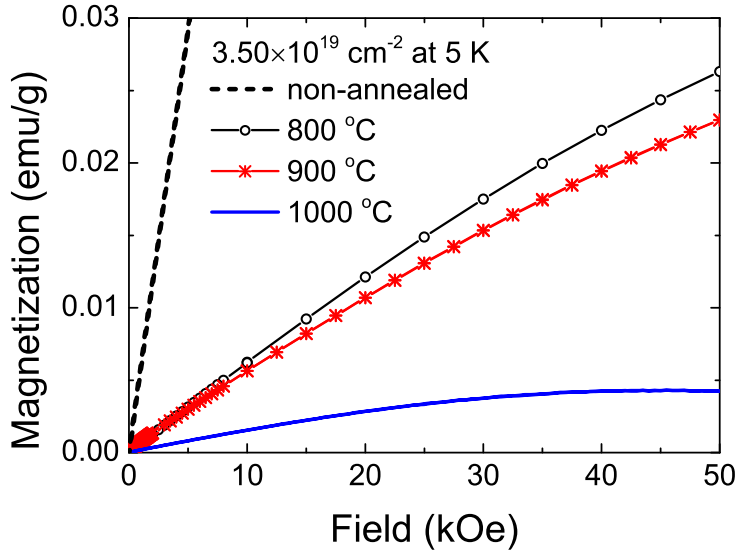}
\caption{Magnetization of irradiated SiC after annealing at different temperature: measurements at
5 K as a function of field [\emph{(M(H)}].} \label{FigMH_annealing}
\end{figure}

Figure \ref{FigMH_annealing} depicts the magnetization of irradiated SiC after thermal annealing at different temperature. As shown in the Raman investigation (Fig. \ref{Raman_annealing}), thermal annealing heals partially the crystalline damage. After annealing, although paramagnetism is still the major component in the magnetic signal, the magnetization is much reduced. The removal of point defects results in the reduction of paramagnetism. The appearance of the ferromagnetic coupling after annealing will be shown in the next subsection.

\begin{figure} \center
\includegraphics[scale=0.4]{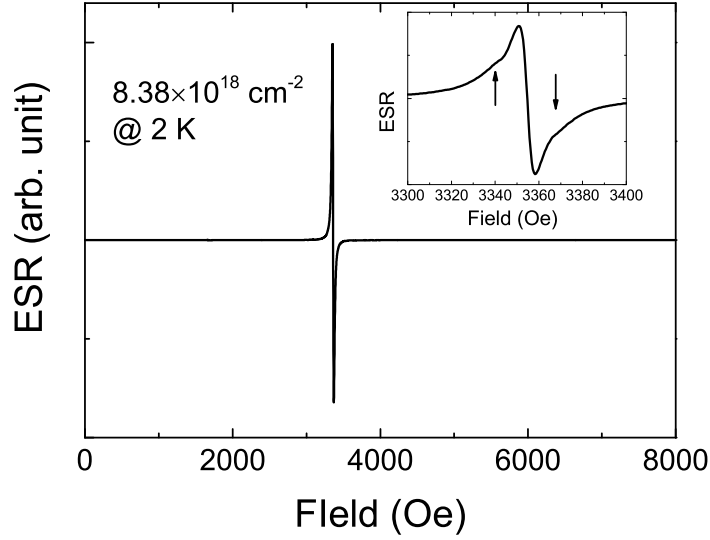}
\caption{ESR spectrum for a 4H-SiC sample irradiated with neutrons with a fluence of 8.38$\times$10$^{18}$ cm$^{-2}$. The field is applied perpendicular to the \emph{c}-axis. The inset shows the detailed hyperfine structure (indicated by arrows) of the resonance peak.} \label{FigESR_nSiC}
\end{figure}

Paramagnetic centers in the neutron irradiated SiC have also been detected by ESR. As shown in Figure \ref{FigESR_nSiC}, in the broad field range, there is only one sharp resonance peak due to paramagnetic electrons. Within the detection limit, there is no broad resonance peak, which could be related with ferromagnetic resonance. The inset shows a narrow scan with smaller field step at the paramagnetic resonance peak. The ESR spectrum exhibits a central line and some weak hyperfine lines. The \emph{g} factor is calculated to be around 2.005, which is the characteristic of free electrons. From the line shape, the defect is very probably the divacancy in 4H-SiC \cite{son2006divacancy}. The hyperfine structure is not well resolved as in the work by Son et al. \cite{son2006divacancy}, which is probably due to the crystal degradation upon neutron irradiation.

\subsection{Ferromagnetism}

\begin{figure*}
\center
\includegraphics[scale=0.5]{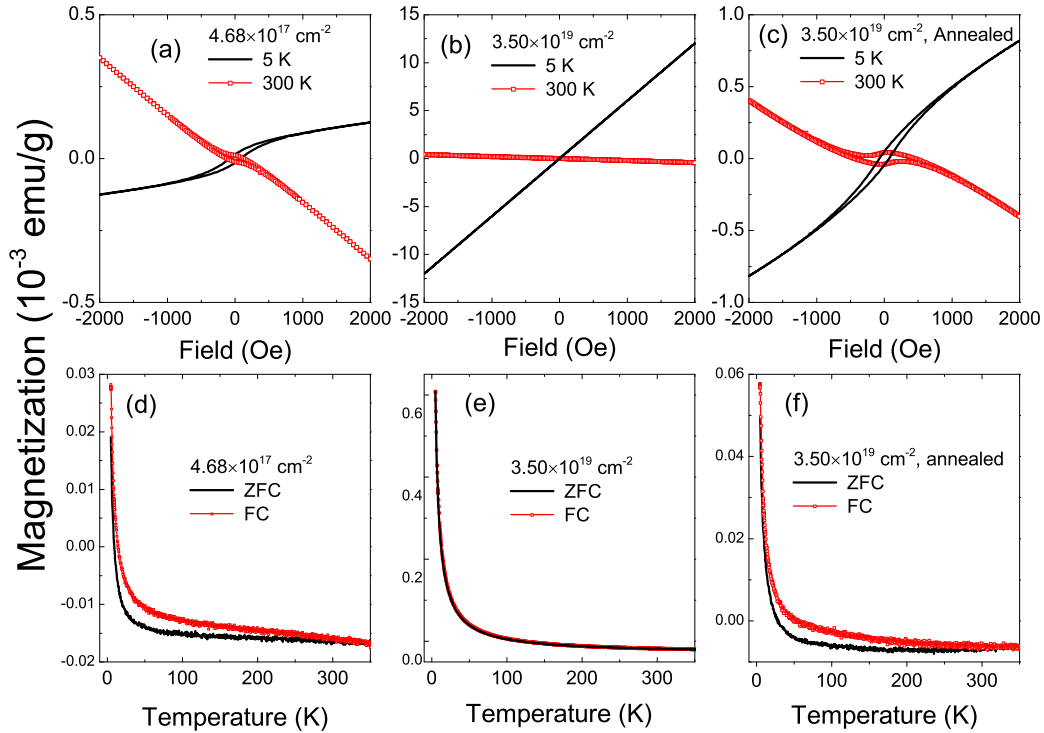}
\caption{Magnetization vs. field at the low field range for samples with fluences (a) 4.68$\times$10$^{17}$ cm$^{-2}$, (b) 3.50$\times$10$^{19}$ cm$^{-2}$ and (c) 3.50$\times$10$^{19}$ cm$^{-2}$ after annealing and ZFC/FC magnetization for the same set of samples (d-f).}
\label{figZFCFC}
\end{figure*}

In Refs. \cite{Liu2011,Lilin}, a sizeable
ferromagnetism was observed for neutron or ion irradiated SiC.
Note that in Ref. \cite{Liu2011}, the maximum
neutron fluence (if counting only the fast neutrons) is around
5.60$\times$10$^{17}$ cm$^{-2}$. The authors show
magnetization measurements at low field and a magnetic hysteresis
loop was observed. We also checked our samples carefully whether a
hysteresis appears in the low field range. Interestingly, besides
the large paramagnetic component we also observe a ferromagnetic
hysteresis in samples with fluence of around 5.60$\times$10$^{17}$
cm$^{-2}$. However, the hysteresis is not resolvable when the
fluence is higher than 6--7$\times$10$^{17}$ cm$^{-2}$. In Fig.
\ref{figZFCFC}, we show the detailed magnetization measurements for
three representative samples: sample 431-50 with an intermediate
fluence 4.68$\times$10$^{17}$ cm$^{-2}$, and sample 431-56 with the
highest fluence of 3.50$\times$10$^{19}$ cm$^{-2}$ before and after
annealing. Sample 431-50 shows a clear hysteresis added on the
paramagnetic component (at 5 K) or on the diamagnetic background
(300 K). In contrast, for sample 431-56, at both 5 and 300 K,
there is no hysteresis resolvable and the paramagnetism dominates
at low temperature. However, after annealing at 900 $^{\circ}$C
for 15 min, sample 431-56 shows a ferromagnetic component at 5 and
300 K. The paramagnetic component is drastically reduced to below
10\% compared with the non-annealed sample due to the reduction of defects as confirmed by Raman spectroscopy (Fig. \ref{FigRaman_nSiC}). Moreover, we also
observed a similar ferromagnetic component in several samples
with the fluence of the order of 2--6$\times$10$^{17}$ cm$^{-2}$ as well as in 6\emph{H}-SiC samples with a large fluence of 3.50$\times$10$^{19}$ cm$^{-2}$) after annealing. The
saturation magnetization for the ferromagnetic component in different
samples is in the range of 1--5$\times$10$^{-4}$ emu/g and is only around 1\% of the paramagnetic component.

\begin{figure} \center
\includegraphics[scale=0.4]{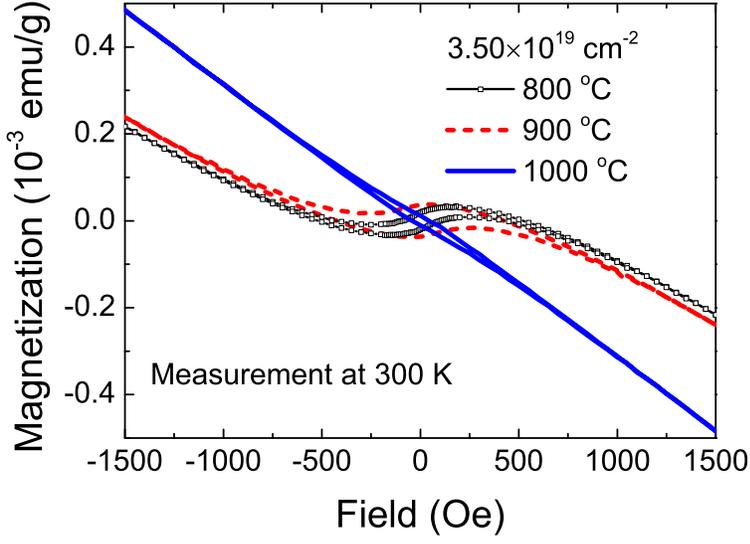}
\caption{Magnetization vs. field at the low field range for
samples after annealing at different temperature. Neither the
diamagnetic background nor the paramagnetic component has been
subtracted from the measured signal.} \label{FM_annealing}
\end{figure}

The appearance or disappearance of the weak ferromagnetic component is confirmed by zero-field-cooled and field-cooled magnetization measurement (ZFC/FC) shown in Fig. \ref{figZFCFC}(d, e, f). The ZFC magnetization was measured by cooling the sample from 350 K to 5 K with zero field, then a field of 100 Oe was applied and the magnetization was measured
during warming up. The FC curves were measured by cooling the sample in a field of 100 Oe during cooling. This approach is often used to verify if the measured specimen is ferromagnetic, paramagnetic or superparamagnetic. If the specimen contains a ferromagnetic component and if the field during measurement is smaller than its coercive field, the ZFC/FC magnetizations will show an irreversibility with the FC magnetization larger than the ZFC magnetization. If the specimen contains only paramagnetism, the ZFC/FC magnetization should overlap with each other. For samples that contain superparamagnetic components, a blocking behavior at low temperature, i.e. an increase in the ZFC magnetization with temperature, will appear. For samples 431-50 [Fig. \ref{figZFCFC}(d)] and 431-56 after annealing [Fig. \ref{figZFCFC}(f)], a difference is observed up to 300 K in the ZFC/FC magnetization, while for sample 431-56 the FC curve superimposes the ZFC curve. Moreover, the data did not indicate any blocking temperatures that can be associated with superparamagnetic behavior, which was reported for ion implanted SiC \cite{wang2014disentangling}. Figure \ref{FM_annealing} shows the magnetization vs. field at the low field range for samples after annealing at different temperature. Annealing at relatively low temperature (in contrast to 1450 $^{\circ}$C normally used for SiC \cite{bohn1987recrystallization}) leads to the appearance of ferromagnetic hysteresis. However, the ferromagnetic component is clearly suppressed when the annealing temperature is increased to 1000 $^{\circ}$C. As shown in Fig. \ref{FigMH_annealing}, the paramagnetism in the sample annealed at 1000 $^{\circ}$C is much reduced in comparison with samples annealed at 800 or 900 $^{\circ}$C. Apparently, annealing at temperatures higher than 900 $^{\circ}$C leads to the annihilation of both isolated and agglomerated divacancies. Moreover, a similar annealing behavior (the reduction of paramagnetism and the appearance of a weak ferromagnetic component) has been observed in 4H-SiC with lower fluences as well as in 6H-SiC (not shown).

\begin{figure}
\center
\includegraphics[scale=0.5]{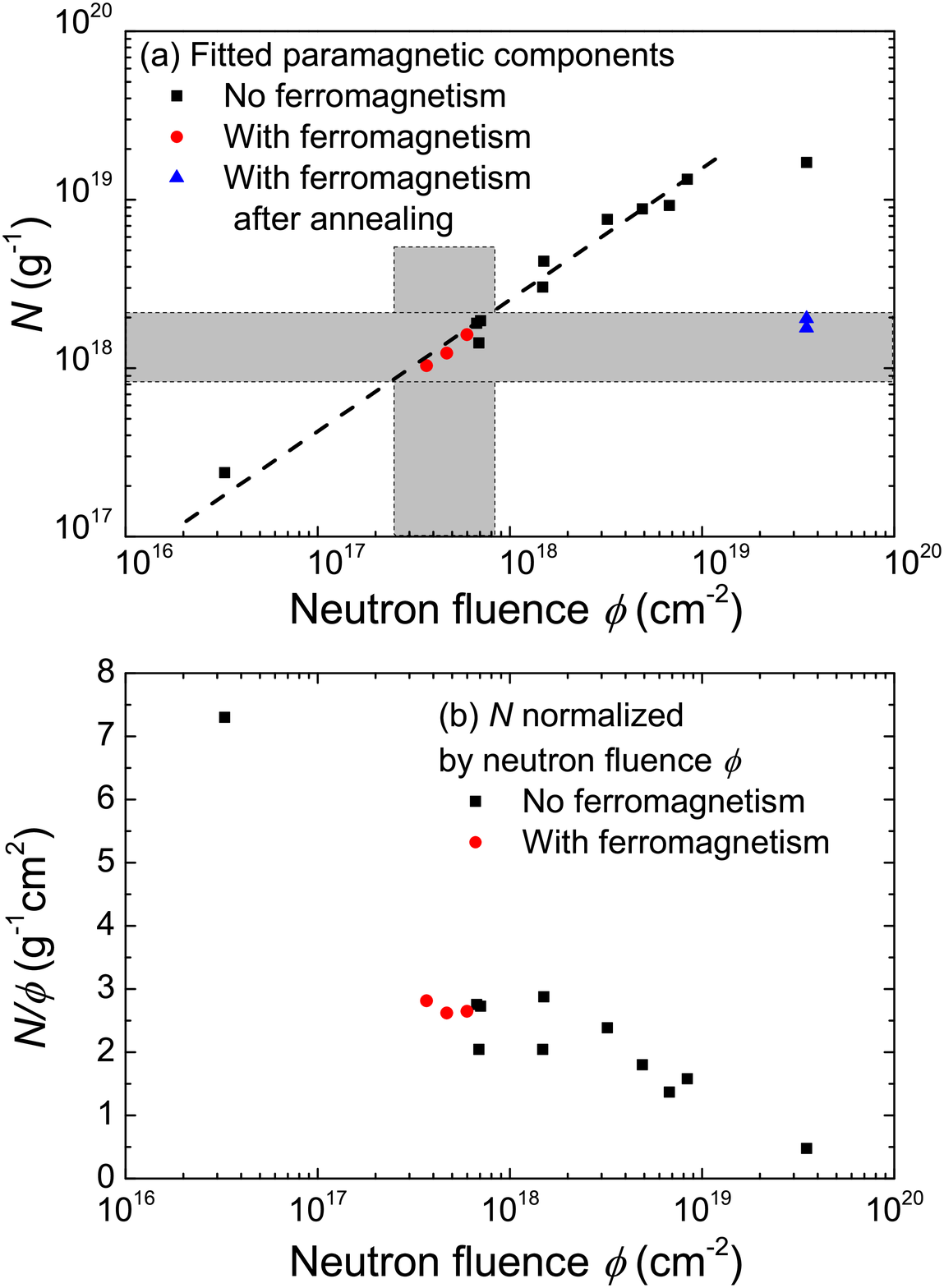}
\caption{(a) The fitted \emph{N} (the density of paramagnetic centers) for samples with different neutron fluence ($\phi$) and the sample with the highest fluence after annealing at 900 $^{\circ}$C for 15 min. The dashed line is only for guiding eyes to indicate the trend. The grey bars indicate the range of samples with a resolvable ferromagnetic component, regarding to \emph{N} and neutron fluence, respectively. (b) \emph{N/$\phi$} vs. neutron fluence. }
\label{figMvsFluence_nSiC}
\end{figure}

As shown by our comprehensive investigation on the magnetic properties of 4H-SiC upon neutron irradiation with a much broader fluence range than that used in Ref. \cite{Liu2011} (especially in the large fluence range), the paramagnetic component scales up with fluence (the concentration of defects), but the weak ferromagnetic contribution appears only for samples with relatively low fluence or after annealing. Figure \ref{figMvsFluence_nSiC}(a) shows the fitted paramagnetism depending on neutron fluence. Those samples denoted by the red circles and by the blue triangle appear to have a ferromagnetic component. As shown in Refs. \cite{Liu2011,Lilin}, the divacancies in SiC can explain those magnetic properties. The scaling between the concentration of paramagnetic centers and the neutron fluence provides solid evidence that the defects created by irradiation are directly responsible for the local moments. However, the long-range-coupling between the local moments is not always favorable. The ferromagnetism only appears when the defect concentration (represented by the density of paramagnetic center and by the neutron fluence) is in a narrow window (as indicated by the grey bars in Fig. \ref{figMvsFluence_nSiC}(a).) This is consistent with the ferromagnetism observed in ion implanted SiC \cite{Lilin,42:/content/aip/journal/jap/115/17/10.1063/1.4860659}. Nevertheless, the saturation magnetization for the ferromagnetic component in different
samples is much smaller than the paramagnetic component. In the next section, we employ first-principles calculations to understand why the defect-induced ferromagnetism is weak.

\begin{figure}
\center
\includegraphics[scale=0.3]{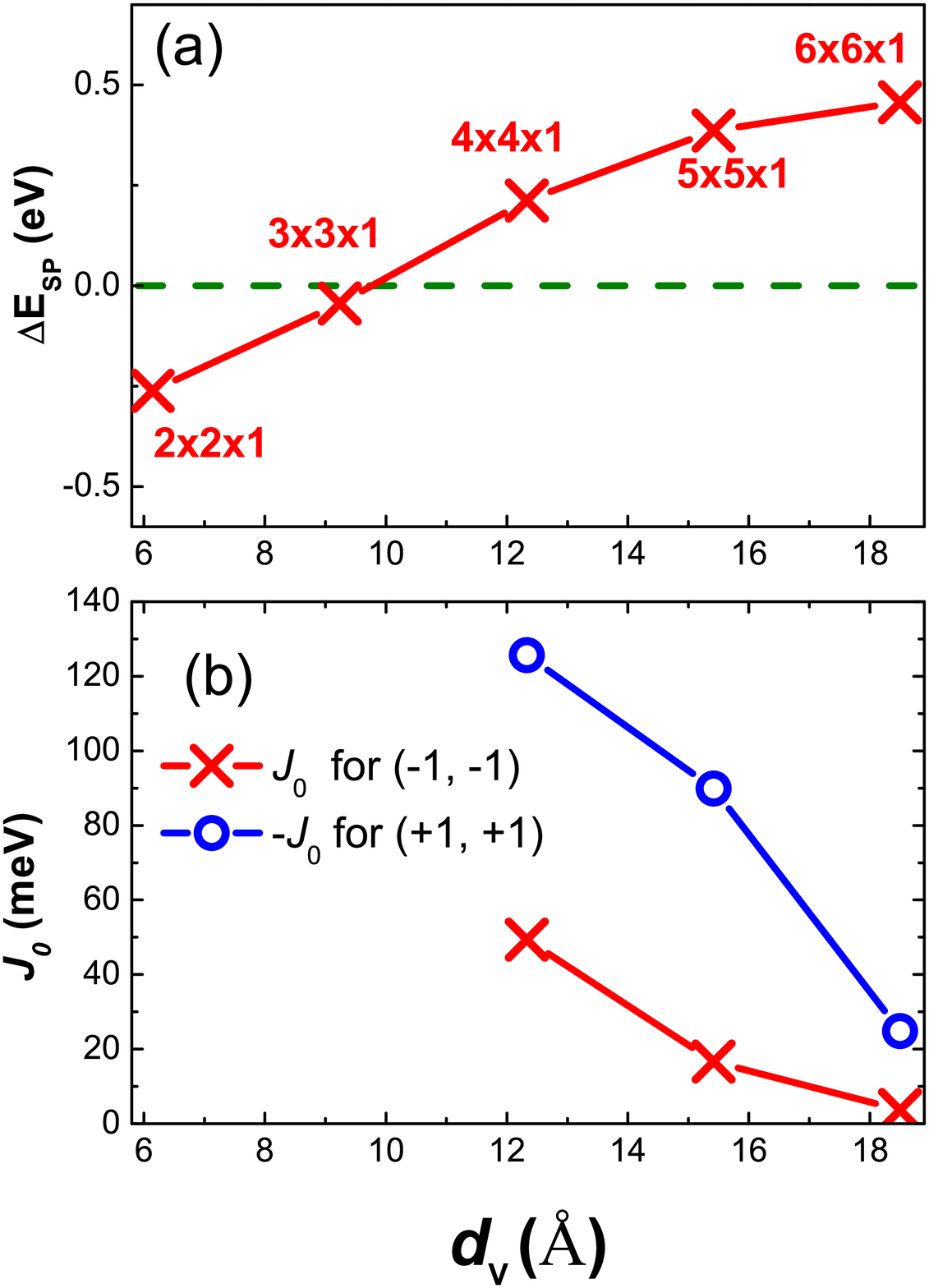}
\caption{(a) Spin-polarization energy $\Delta$E$_{sp}$ as a function of the distance \emph{d$_{V}$} between the adjacent vacancies. $\Delta$E$_{sp}$ increases with increasing \emph{d$_{V}$} and the critical value is around 10 \AA. (b) The ferromagnetic (or antiferromagnetic) exchange coupling energy between charged (-1, -1) or (+1, +1) divacancies as a function of \emph{d$_{V}$}. The corresponding supercell structures with one V$_{Si}$V$_{C}$ are marked with the numbers and multiply signs.}
\label{SiC_neutron_DFT}
\end{figure}

\subsection{Why is the ferromagnetic signal weak?}

\begin{table}
\begin{center}
\caption{\label{tab:sample} Magnetic coupling between
V$_{Si}$V$_{C}$-induced local moments in SiC. The coupling is
antiferromagnetic between neutral or positively charged
V$_{Si}$V$_{C}$ while ferromagnetic between negatively charged
ones. As the arrangement direction of V$_{Si}$V$_{C}$ is along the
\emph{c} axis, the coupling almost does not exist.}
\begin{tabular}{ccccc}
\hline\hline
  \emph{d$_V$} (\AA) & \emph{c}- or \emph{a}-axis & charge state & E$_{AFM}$-E$_{FM}$ (meV) & J$_0$ (meV) \\
  \hline
 10.11 & \emph{c} axis & (0, 0) & -0.55 & -0.14 \\
 12.33 & \emph{a} axis & (0, 0) & -39.45 & -4.93 \\
 12.33 & \emph{a} axis & (+1 +1) & -251.35  & -125.68 \\
 12.33 & \emph{a} axis & (-1, -1) & 98.73 & 49.36 \\
 15.42 & \emph{a} axis & (0, 0) & -5.77 & -0.72 \\
 15.42 & \emph{a} axis & (+1, +1) & -179.81 & -89.90 \\
 15.42 & \emph{a} axis & (-1, -1)& 33.31 & 16.66 \\
 18.50 & \emph{a} axis & (0, 0) & -0.93 & -0.12 \\
 18.50 & \emph{a} axis & (+1, +1) & -49.75 & -24.88 \\
 18.50 & \emph{a} axis & (-1, -1) & 7.43 & 3.71 \\
 %12.34 & 6H & \emph{a} axis & -37.48 (0, 0) & -4.68 \\
 %12.34 & 6H & \emph{a} axis & 98.50(-1 -1) & 49.25 \\
 %15.14 & 6H & \emph{c} axis & 0.024 (0, 0) & 0.006 \\
 %15.42 & 6H & \emph{a} axis & -6.67 (0, 0) & -0.83 \\
 \hline\hline
 \end{tabular}
 \end{center}
\end{table}

According to our previous positron annihilation experiments
\cite{Liu2011,Lilin}, divacancies have been identified in neutron or ion irradiated SiC and are the most likely origin for the measured ferromagnetism. The ferromagnetism only occurs in samples with relatively low neutron
fluence or after annealing and is much weaker compared with the induced paramagnetism. These facts motivated us to perform some theory work to understand the puzzle. First-principles
calculations were performed by using the Cambridge Serial Total Energy
Package \cite{clark2005first}. Spin-polarized electronic structure
calculations were carried out using the Perdew-Burke-Ernzerhof
functional \cite{GGA} for the
exchange-correlation potential based on the generalized gradient
approximation. The core-valence interaction was described by
ultrasoft pseudopotentials \cite{ISI:A1990CZ84300072}. The cutoff
energy for the plane-wave basis is set to 310 eV. Full
optimization of the atomic positions and lattice parameters was
carried out with convergence threshold of the remanent
Hellmann-Feynman force 0.01 eV/\AA. 4H-SiC supercells with various
sizes containing one axial divacancy (V$_{Si}$V$_{C}$)
of built to obtain the variation of spin polarization and the
magnetic coupling depending on the distance between defects. Actually, the calculation results show no significant difference between 4H- and 6H-SiC.

As a hexagonal structure, 4H-SiC has a lattice constant \emph{c} (out-of-plane) different from \emph{a} and \emph{b} (in-plane), so the distance between adjacent V$_{Si}$V$_{C}$ divacancies along different axes is also different from each other. Usually, the magnetic moments couple with each other preferentially along the shortest distance, which will be chosen as the coupling distance. It is found that the coupling is along the \emph{a-b} plane in a SiC supercell, which will be explained hereinafter. Therefore, the distance between adjacent V$_{Si}$V$_{C}$ in the \emph{a-b} plane is chosen as the coupling distance (\emph{d$_{V}$}) between them. The spin-polarizing energies $\Delta$E$_{sp}$ (the energy difference between the spin-unpolarized and spin-polarized states) of supercells as a function of \emph{d$_{V}$} are shown in Fig. \ref{SiC_neutron_DFT}. It is found that $\Delta$E$_{sp}$ increases when \emph{d$_{V}$} increases, indicating that the spin polarization is stable under low V$_{Si}$V$_{C}$ concentrations. When \emph{d$_{V}$} is larger than around 10 \AA, $\Delta$E$_{sp}$ changes from negative to positive and the spin polarization becomes energetically favored. Here, the supercells with \emph{d$_{V}$} less than around 10 \AA~are designed for obtaining the critical value of \emph{d$_{V}$} switching the sign of $\Delta$E$_{sp}$. Besides, each V$_{Si}$V$_{C}$ yields a local moment of 2.0 $\mu_B$ when \emph{d$_{V}$} is more than 12.34 \AA. This result is in good agreement with our experimental data shown in Figs. \ref{figMH_nSiC} and \ref{FigMT_nSiC}. The experimental results show that the induced paramagnetism is due to a single species with a moment of 2 $\mu_B$. Moreover, as shown in Fig. \ref{figMvsFluence_nSiC}(b), the normalized density of paramagnetic center by the neutron fluence (i.e. the measure of the amount of defects), $N/\phi$ decreases with the neutron fluence. This trend again indicates that a low defect-density favors spin-polarization as shown in Fig. \ref{SiC_neutron_DFT}(a).

In order to investigate the magnetic coupling (ferromagnetism or antiferromagnetism) between these V$_{Si}$V$_{C}$-induced local moments, we put two 4H-SiC 4$\times$4$\times$1 supercells side by side along the \emph{a}- or along the \emph{c}- direction. Each 4$\times$4$\times$1 supercell contains one V$_{Si}$V$_{C}$. Note that these antiferromagnetic structures are designed only for obtaining the magnetic interaction. The energy difference between the antiferromagnetic and ferromagnetic phases is
$E_{AFM}-E_{FM}=8J_{0}(d_{V})S^{2}$ or $E_{AFM}-E_{FM}=4J_{0}(d_{V})S^{2}$ for size doubling along \emph{a} or \emph{c} directions, respectively, according to the nearest-neighbor Heisenberg model, where $J_0(d_V)$ is the nearest-neighbor exchange interaction as a function of \emph{d$_V$} and \emph{S} is the net spin of the V$_{Si}$V$_{C}$ states. In previous work, it has been shown that the intrinsic defects in high purity SiC can be neutral, positively or negatively charged \cite{janzen2006intrinsic}. The possible impurities, such as Boron and Nitrogen (see ref. \onlinecite{ISI:000182770100028}), with the concentration of the order of 10$^{15-16}$ cm$^{-3}$ can also modify the charge state of V$_{Si}$V$_{C}$. Therefore, we calculate the interaction for different charge states. As shown in Table I, V$_{Si}$V$_{C}$ does not couple with each other along the \emph{c} axis, even when \emph{d$_V$} is as small as 10.11 \AA. The exchange interaction is much stronger along the \emph{a} axis and strongly depends on the charge state. Neutron or positively charged V$_{Si}$V$_{C}$ antiferromagnetically couples with a maximal exchange interaction of 125.68 meV when \emph{d$_V$} is 12.33 \AA. The result is also consistent with the previous report \cite{Liu2011}, in which the coupling between neutral V$_{Si}$V$_{C}$ is also antiferromagnetic. Negatively charged V$_{Si}$V$_{C}$ divacancies favor a ferromagnetic coupling. However, the exchange interaction dramatically decreases from 49.36 meV to 3.71 meV then \emph{d$_V$} is increased from 12.33 to 18.50 \AA.

The mutual dependence of spin-polarization and exchange interaction on the concentration of defects is not only suggested for SiC, but also for graphene/graphite \cite{yazyev2007defect,palacios2008vacancy} as well as for nitrides \cite{dev:117204,liu2012adjustable} and oxides \cite{pemmaraju2005ferromagnetism}. To develop the long-range ferromagnetic coupling, it is also crucial to keep the stacking order in graphite \cite{yazyev2008magnetism}.

In Figure \ref{SiC_neutron_DFT}(b), we plot the ferromagnetic exchange interaction energy depending on \emph{d$_V$}. As shown in Fig. \ref{SiC_neutron_DFT}(a) and (b), the ferromagnetism induced by divacancies requires that divacancies are negatively charged with \emph{d$_V$} in the range from 12.33 \AA~to 15.42 \AA. The required \emph{d$_V$} would correspond to a concentration of divacancies of the order of 3--6$\times$10$^{20}$ cm$^{-3}$ (1/\emph{d$_V^3$}). This concentration is around 1 at.\% of SiC and well above the maximum concentration for the local moments we obtained by neutron irradiation (see Fig. \ref{figMvsFluence_nSiC}). However, after irradiation with the largest neutron fluence the density of paramagnetic centers (\emph{N}) starts to saturate probably due to the appearance of local regions which contains a large density of defects and tends to be non-spinpolarized as shown in Fig. \ref{SiC_neutron_DFT}(a). Moreover, it is also crucial to keep the crystalline ordering for long-range ferromagnetic coupling \cite{yazyev2008magnetism}. In this sense, it is unrealistic to expect ferromagnetic coupling throughout the whole bulk sample. In principle, it is in agreement with our experimental observation. The experimentally observed weak ferromagnetism probably can be understood as a local effect: only some particular regions with nm or $\mu$m dimension can accommodate a high concentration of divacancies. These regions form ferromagnetic bubbles with very strong interaction (see Table I), leading to the high Curie temperature. Presumably, these regions could be at the surface, an interface or a grain boundary. However, according to our calculation the spin-polarization will be unfavorable if \emph{d$_V$} is too small. Therefore, the narrow fluence window of ferromagnetism shown in Fig. \ref{figMvsFluence_nSiC} as well as in neon implanted SiC \cite{Lilin} can be speculated: In the beginning of neutron irradiation, the defect concentration increases, and some local regions accommodating a larger concentration of divacancies appear to be ferromagnetic. When \emph{d$_V$} in the local regions reaches the critical value of about 10 \AA, the increase of the defect concentration will suppress the spin polarization. There may appear some new local regions with the concentration of divacancies reaching to the level to induce ferromagnetism, but the damage to the crystalline structure by irradiation will weaken the coupling. Also, the existence of minor and major carriers in SiC can lead anti-ferromagnetic coupling and cancel the ferromagnetic coupling to some degree. Therefore, the magnetic moments can not be increased by increasing the fluence of irradiation all the time, while, the amount of paramagnetic centers can be scaled up with neutron fluence and show saturation at very large fluences. The locally accumulation of ferromagnetic defects has also been suggested for defective ferromagnetic oxides \cite{straumal2009magnetization,Jcoey1367,straumal2013grain}.

To clarify if the defect induced ferromagnetism is associated with some local regions, it rather calls for a sophisticated investigation using methods which are sensitive to the magnetic order and provide an enough spatial resolution at the same time. Spin polarized scanning tunneling microscopy \cite{Ugeda2010a} or muon spin relaxation \cite{storchak2008spatially,ricco2011muons} could provide us a further insight, which is beyond the scope of the present paper.

%As neutrons have strong penetration ability, the domains of defects will also expand along c axis, which can not make the magnetism relatively strong, which should be the reason for the relative weak magnetization in SiC after neutron irradiation. This a-b plane energy favor model is also consistent with the previous XTEM research of defect evolution under annealing. It pointed that defect evolve into the two-dimensional cluster when vacancy migration begin. In this process, single defect should be wiped out below 700 $^{\circ}$C and then more and more divacancies involve into the two-dimensional cluster \cite{barbot2009helium,leclerc2008evolution}.

\section{Conclusion}

To investigate defect-induced magnetism in SiC, we applied a broad neutron fluence covering four orders of magnitude to generate defects in a controlled way. A huge paramagnetic contribution is observed and scaled up with neutron fluence, whereas ferromagnetism only appears in a certain, rather low fluence range or after annealing treatments. First-principles calculations hint towards a mutually exclusive role of defects: A low defect concentration favors spin-polarization, but leads to negligible magnetic interaction. Combining both experimental and first-principles calculation results, we conclude that defect-induced ferromagnetism cannot be scaled up with the volume. It should be rather understood as local effect. Therefore, our investigation answers why the measured magnetization is low for defect-induced ferromagnetism in SiC.

Financial support from the Helmholtz-Gemeinschaft Deutscher
Forschungszentren (VH-NG-713, VH-VI-442 and PD-146) is gratefully acknowledged. The authors also acknowledge the support by the International Science and Technology Cooperation Program of China (2012DFA51430).

%\bibliography{../thesis}

\end{document}